\documentclass{osa-article}
\pdfoutput=1

\journal{osajournal}

\articletype{Research Article}

\usepackage{lineno}
\usepackage{physics}

\begin{document}

\title{Measurement of absolute radius, refractive index and dispersion of a long cylinder}

\author{Mathieu Couillard,\authormark{1,2,*} and Pablo Bianucci,\authormark{1}}

\address{\authormark{1} Physics Department, Concordia University, 7141 Sherbrooke Street W, Montreal, Quebec H4B 1R6, Canada

\authormark{2} Experimental Quantum Information Physics Unit, Okinawa Institute of Science and Technology, 1919-1 Tancha, Onna, Kunigami, Okinawa, 904-0495, Japan
}

\email{\authormark{*}mathieu.couillard@oist.jp} 


\newcommand{\pbchange}[1]{{\color{red}{#1}}}
\newcommand{\pbcomment}[1]{{\color{blue}{\{#1\}}}}


\begin{abstract}
Long cylinders, such as optical fibers, are some of the most widely used photonic devices. The radius and refractive index of these fibers are therefore fundamentally important parameters in determining their performance. We have developed a method to determine the absolute radius, refractive index, and chromatic dispersion of a long cylinder using only the resonance wavelengths of the whispering gallery modes around its circumference for two different polarizations. Since this method only requires the measurement of resonance wavelengths, it is non-destructive and it can be performed using standard equipment.  As a proof-of-concept, we demonstrate the method on a $125\mu m$ optical fiber and an $80\mu m$ borosilicate capillary fiber with thick walls, obtaining values for the diameter and the refractive index with an accuracy of $2 nm$ and $2\times 10^{-5}$, respectively.
\end{abstract}

\section{Introduction}
Optical fibers are one of the pillars of modern photonics technology\cite{fang2012fundamentals}. The accurate determination of the diameter, dispersion, and refractive index of an optical fiber is important for applications such as fiber-based sensors\cite{s140405823}, and controlling the propagation of light for telecommunications\cite{agrawal2012fiber}. As optical fibers have a circular cross-section, they can sustain whispering gallery modes (WGMs). In a WGM, light continuously undergoes total internal reflection on the surface of a circular transparent material surrounded by a one with a lower refractive index.  Due to the low loss of typically used transparent materials, and the very smooth surfaces that can be obtained, these modes are capable of reaching ultrahigh Q factors greater that $10^{10}$~\cite{savchenkov2004kilohertz}. The achievable ultrahigh quality factors have made whispering gallery resonators (WGRs) excellent candidates for sensing~\cite{yang2016high, dong2009fabrication, bogatyrev2002change, liang2017resonant, vollmer2002protein, ksendzov2004integrated, jiang2020whispering}and for observing non-linear effects~\cite{del2007optical, arnold2009whispering, spillane2002ultralow, grudinin2009brillouin, del2007optical}.

When characterizing optical fibers and WGRs, the refractive index is crucial in determining their optical behavior. There are already precise methods used to measure the refractive index like refractive near-field scanning\cite{stewart1982optical} but, they require a dedicated setup and are generally destructive. The refractive index of a material can also be calculated using an empirical formulation like the Sellmeier equation\cite{ghosh1997sellmeier}, but given the plethora of glasses and dopants used by manufacturers (whose composition is generally treated as a trade secret) it may not be possible to find an appropriate formula. It has also been shown that by finding changes of the resonance wavelengths of WGMs, it is possible to determine variations of the effective radius of an optical fiber with high precision (the effective radius is defined as the product between the fiber radius and its refractive index)\cite{birks2000high,sumetsky2010radius}. Shortly after, it has also been shown that WGM can be used to determine the size and refractive index of a microspheres\cite{preston2015determining} by precisely matching the modes to theoretical values. In this work, we reach a similar goal for a long cylinder using a method that is significantly simpler both in determining the mode numbers and minimizing the error. We determine all the mode numbers, radius, refractive index, and chromatic dispersion of a long cylinder by only using data from its mode spectra for two orthogonal polarizations and the assumption that the WGR has insignificant deviations from a circular cross-section.

We start by describing the experimental setup, followed by methods to calculate values for the radial and azimuthal mode numbers given the polarization and approximate values for the radius and the refractive index. From this initial mode identification, we then vary parameters such that the radius measured from each mode gives the same result. As an example, we demonstrate the method on two fibers of different sizes and materials.

\section{Experimental setup}
\begin{figure}[htbp]
	\centering\includegraphics[width=7cm]{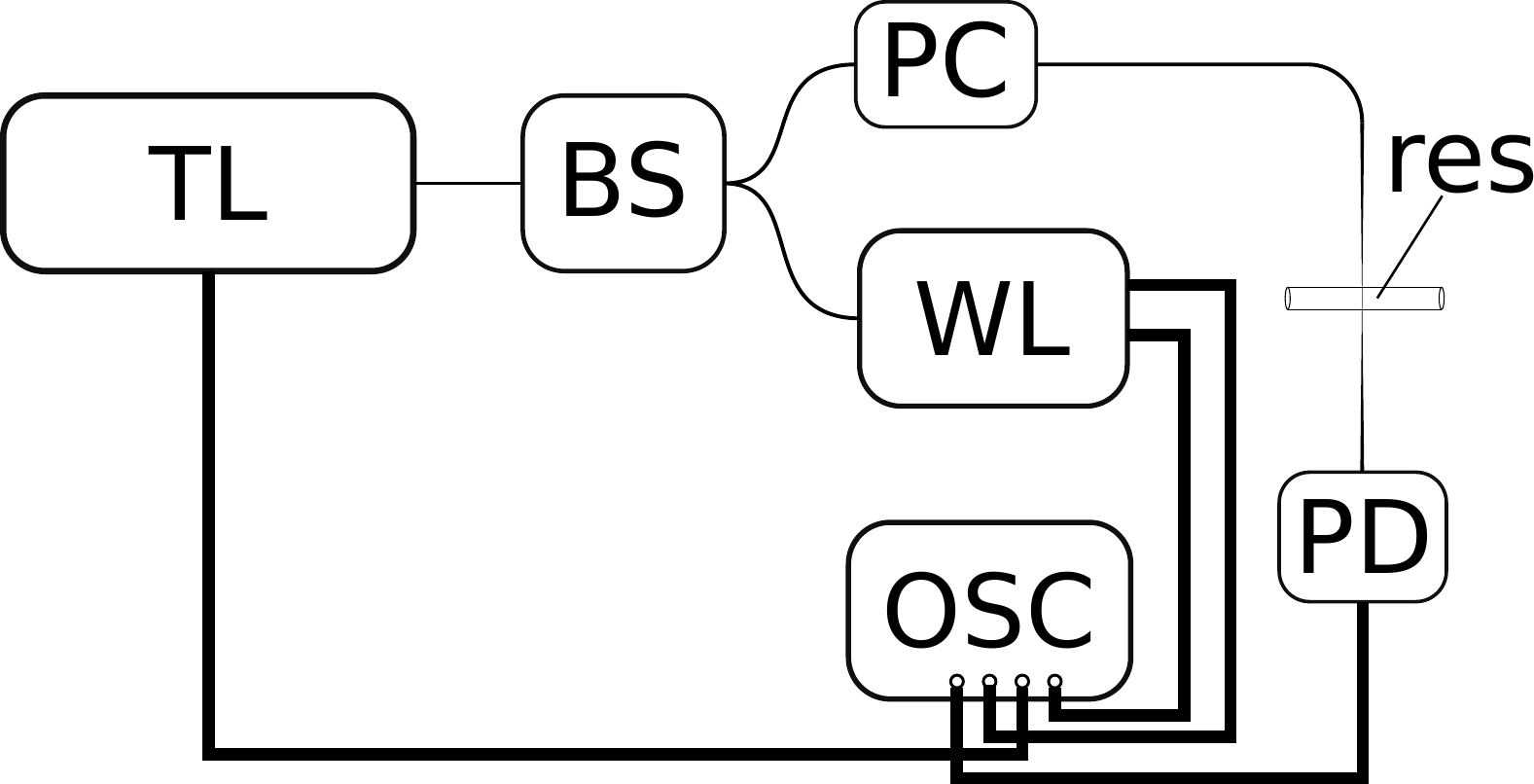}
	\caption{The experimental setup. Thick lines represent electrical wires and thin lines represent optical fibers. TL: Tunable laser, BS: beam splitter, PC: polarization controller, res: whispering gallery resonator, PD:photodetector, WL: wavelength locker, OSC: oscilloscope.}
\label{fig:setup}
\end{figure}

\begin{figure}[htbp]
	\centering\includegraphics[width=7cm]{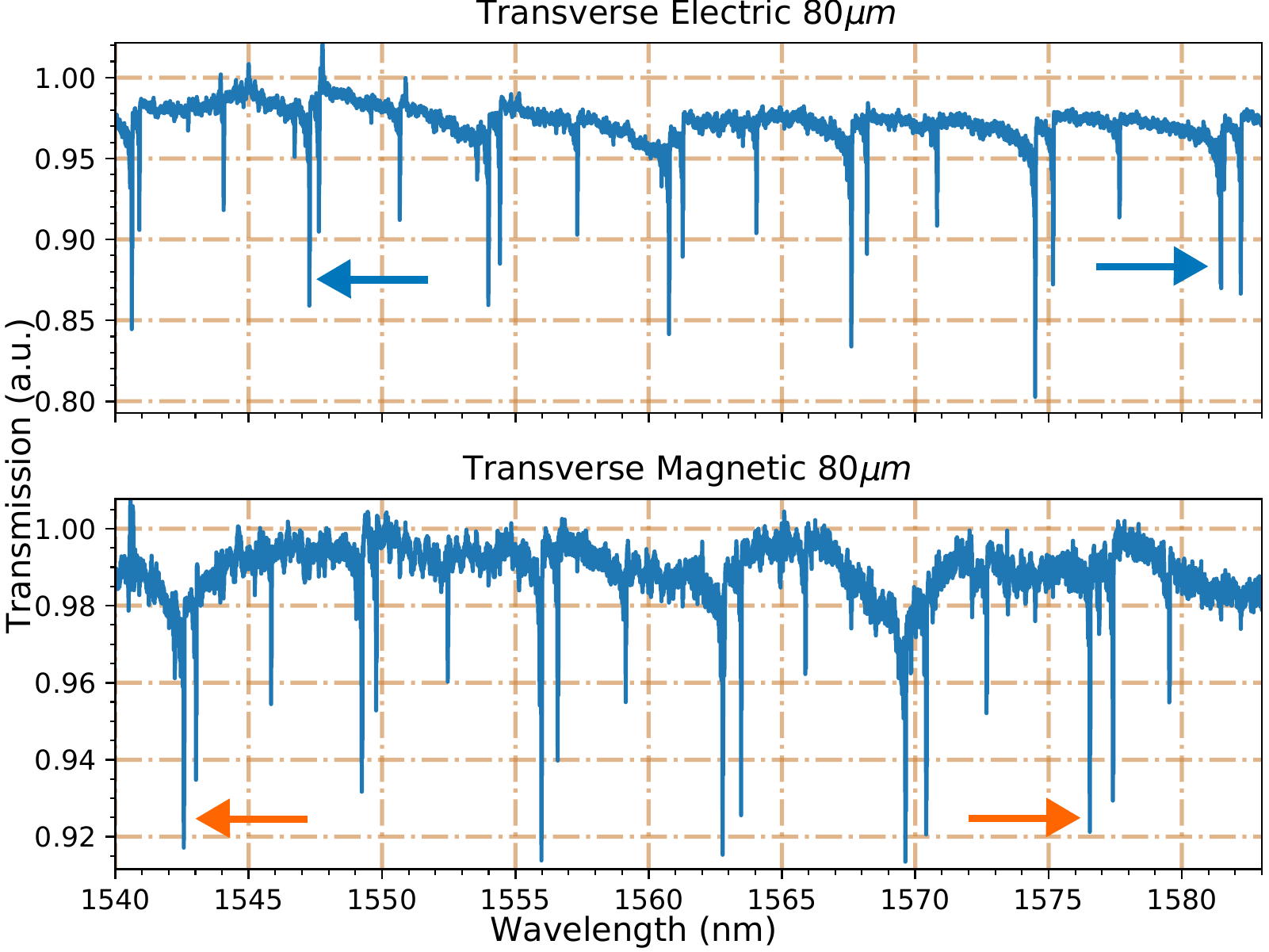}
	\caption{WGM transmission spectra for modes with orthogonal polarizations for a WGR with a diameter of $80\mu m$. The arrows indicate the modes used in the analysis depicted in Fig.~\ref{fig:ratio}.}
\label{fig:spectrum}
\end{figure}

The experimental setup used to collect the WGM spectrum is show in Fig.~\ref{fig:setup}. A tunable laser scans over a range of wavelengths, sending a small portion of the light to a wavelength locker for wavelength measurement. Most of the light is sent to the fiber under test, where it couples in and out of the WGR via a tapered optical fiber, fabricated using the flame brush technique\cite{ward2014contributed}. The transmission, after interacting with the WGR, is sent to a photodiode, and collected in an oscilloscope together with the wavelength locker signal. Typical spectra for both polarizations, collected in this case from a resonator with a diameter of $80\mu m$, are shown in Fig.~\ref{fig:spectrum}. The spectra with different polarizations were measured at the same physical position along the fiber.

\section{Method and results}
Once the spectra are measured, we can extract the resonance wavelength of different WGMs by locating their wavelengths of minimum transmission (i.e. the bottom of the corresponding transmission dip). 
With the resonant wavelengths determined, we can now identify the radial mode number $p$, the azimuthal mode number $m$ and the polarization (transverse electric, TE, or transverse magnetic, TM, where we define the transverse direction to be along the long axis of the fiber). 

To determine the mode numbers and polarization, we use the transcendental equation for a two-dimensional disk\cite{BianucciJAP09},
\begin{equation}
		P\frac{[(nkR)^{1/2}J_m(nkR)]'}{(nkR)^{1/2}J_m(kR)}=\frac{[(kR)^{1/2}H_m^{(1)}(kR)]'}{(kR)^{1/2}H_m^{(1)}(kR)},
		\label{eq:trans}
\end{equation}
where $n$ is the refractive index, and $P=n$ or $1/n$ for TE and TM polarizations, respectively; $k=2\pi/\lambda$ is the free space wavenumber where $\lambda$ corresponds to the resonance wavelength, $R$ is the resonator radius, $J_m(x)$ and $H^{(1)}_m(x)$ are the $m^{th}$-order cylindrical Bessel and Hankel function of first kind, respectively, with the prime denoting a derivative with respect to the argument $nkR$ and $kR$ for the left hand side and right hand side, respectively. The order of the Bessel and Hankel functions corresponds to the azimuthal mode number $m$. Solutions to this equation for fixed parameters correspond to different radial mode numbers $p$ (with the number increasing as the solution increases), and give the dimensionless product $kR$ of the mode. Each resonant WGM, with given polarization and mode numbers, corresponds to a particular solution of Eq.~\ref{eq:trans}.

Identifying $p$ for a resonance can be done by comparing the free spectral ranges (FSRs) of different mode families. For a given value of $p$, the FSR is very well approximated as constant for a small range of $m$. This was used to identify which modes have the same $p$, differing only by $m$. It can easily be shown by Eq.\ref{eq:trans} that modes with smaller $p$ have smaller FSR. Conceptually, this can be understood by the modes with larger $p$ passing closer the center of the fiber, therefore having a smaller effective radius and larger FSR. We can therefore determine the mode families by ordering them by increasing FSR, starting with $p=1$ for the lowest mode family.

To determine $m$, we consider two distinct resonant modes (with the same polarization) $a$ and $b$, with corresponding resonance wavelengths $\lambda_a$ and $\lambda_b$. The ratio between the wavelengths is also equal to the inverse ratio of the corresponding solutions to Eq.~\ref{eq:trans}:
\begin{equation}
	\frac{\lambda_b}{\lambda_a} = \frac{k_aR}{k_bR}.
	\label{eq:ratio}
\end{equation}
Taking estimated values for the fiber refractive index, we can find solutions for Eq.~\ref{eq:trans} for pairs of values $m$ and $m+\Delta m$, where plausible values of $m$ are estimated from $m\approx 2\pi Rn/\lambda$\cite{birks2000high}. The value of ${k_mR}/{k_{m+\Delta m}R}$ for which the ratio is closest to the experimentally measured one is then our ``best fit" for the mode.

We measured the spectra for two different fibers: the first was a Vitrocom CV0508 borosilicate hollow glass tubing with $15\mu m$ walls and a nominal outer diameter of 80 $\mu$m, and the second a nominal $125\mu m$-diameter Corning LEAF optical fiber. Despite the former being hollow, the modes used in this experiment did not have significant spatial overlap with the hollow region of the tube for all the observed $p$, such that  Eq.~\ref{eq:trans} accurately describes them. We use these two fibers to demonstrate that the results are similar for resonators of different size, materials, and manufacturers.

As an example, we choose the resonant wavelengths of the dips marked by arrows in Fig.~\ref{fig:spectrum}, which correspond to modes with $p=3$, where we identified $p$ using the FSR, as described previously, and $\Delta m = 5$ by counting the number of dips between the two chosen modes.

\begin{figure}[htbp]
\centering\includegraphics[width=7cm]{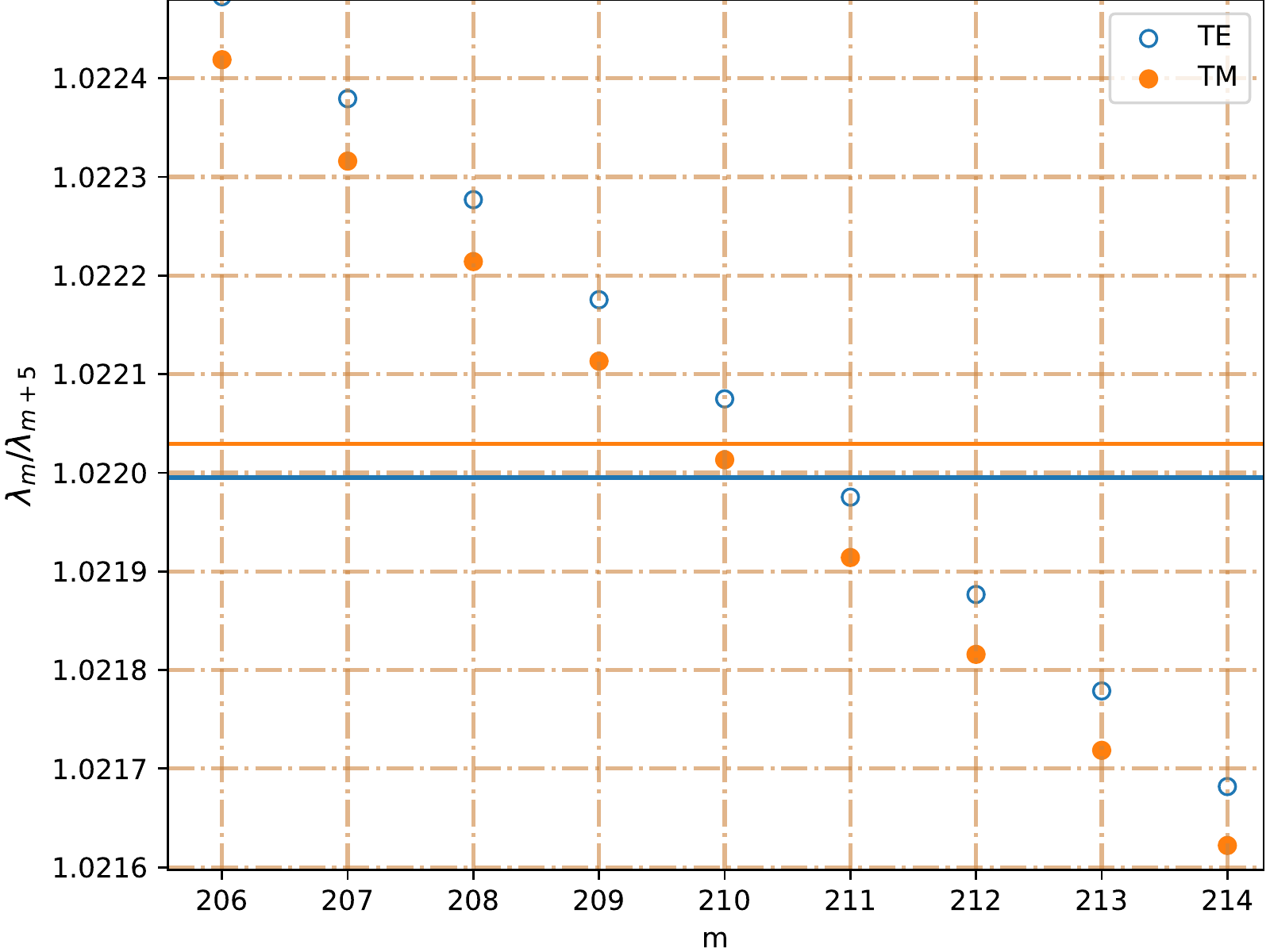}
\caption{Ratio of two resonance wavelengths as a function of mode number. The dots represent the theoretical values of the wavelength ratio of 2 modes as a function of the azimuthal mode number $m$. The modes have $p=3$ and differ by $\Delta m=5$. The horizontal lines mark the experimentally measured ratio.}
\label{fig:ratio}
\end{figure}

Fig.~\ref{fig:ratio} shows the result from calculating the ratio between solutions to the transcendental equation for two modes with $m$ and $m + 5$ as a function of $m$, for both polarizations. The horizontal lines indicate the experimental values for the ratio (Eq.~\eqref{eq:ratio}) and the dots represent each numerically calculated ratio. The points closest to the experimental values then determine the value of $m$  and the polarization (in this case, $m = 221$ for TE and $m = 220$ for TM). The same calculation was then repeated for the other radial modes visible in the spectrum. The initial refractive indices were chosen to be the values given in the corresponding datasheets: 1.473 for the for the nominally $80\mu m$ fibers\cite{schott2019} and 1.4693 for the nominally $125\mu m$ fiber\cite{corning_leaf}. These values are only estimates since they do not correspond to the measured wavelength,  and the quoted value for the $125\mu m$ fiber corresponds to the group index of the core. This means that the values for $m$ we just found are also only estimates, which still need to be corrected.

\medskip

\begin{figure}[htbp]
\centering\includegraphics[width=\textwidth]{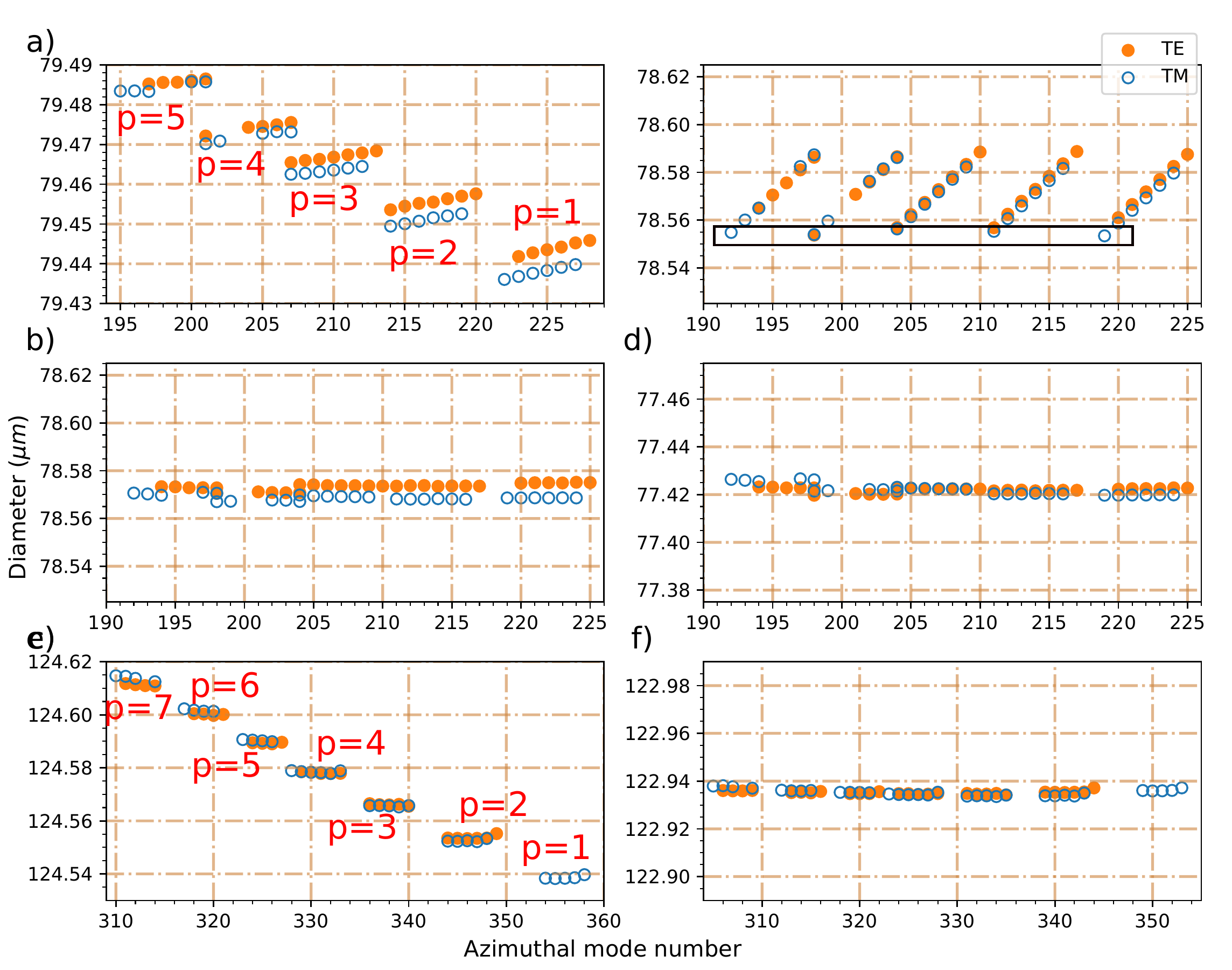}
\caption{Determined diameter as a function of $m$ at different steps of the analysis
process. a), b), c), and d) correspond to the $80 \mu m$ nominal diameter fiber, while e) and f) correspond to the $125 \mu m$ nominal diameter one. a) and e) are the diameters from the initial estimates. b) shows the diameter after the value of $m$ was corrected. c) shows the result after chromatic dispersion is added. d) and f) are the final calculated values. Note that the $p=1$ radial mode for the $125\mu m$ fiber was difficult to observe in the TE polarization due to overcoupling, therefore we omitted these values.}
\label{fig:radii}
\end{figure}

Once all the mode numbers are calculated, we can determine $R$ by first calculating the root of Eq.~\eqref{eq:trans} for the corresponding set of mode numbers and polarization, and then dividing this value by the measured $k=2\pi n/\lambda$. If we initially assume the same $n$ as before and no chromatic dispersion, the measured values of the diameter, which equals twice the radius as a function of $m$ are plotted in Fig.~\ref{fig:radii}a) and e) for the 80 $\mu$m and 125 $\mu$m diameter fibers, respectively.

We see that the estimated values for the diameter differ substantially, as expected given that the refractive index was initially estimated and the chromatic dispersion was neglected. Since the spectra were taken all at the same physical position, we can safely assume that the obtained diameter should be the same for all modes. Thus, we are now tasked with varying the parameters until all the modes result in the same diameter. We first note, as noticeable in Fig~\ref{fig:radii}a), that the results for different modes are grouped by common $p$, and that the measured diameter increases with increasing $p$. This staircase-like structure is a consequence of our estimated $n$ when calculating $m$ in Eq.~\ref{eq:ratio} and our neglect of chromatic dispersion. We can correct this by varying $m$ until modes with similar wavelength and the same polarization agree on the calculated diameter, leading to Fig.\ref{fig:radii}b) where the black box indicates a group of modes with similar wavelength. The increasing slopes seen within each group of radial modes can then be corrected by varying the chromatic dispersion (which we assume to be linear, a valid assumption for a relatively narrow bandwidth away from material resonances) until all the modes of the same polarization agree, as shown in Fig~\ref{fig:radii}c). These values agree well with the typical dispersion of glasses\cite{stephens1954refractive}. At this stage, we have already corrected the $m$ values and included chromatic dispersion, but we have not found the correct value of the base refractive index yet. Observing that the dependence of Eq.~\ref{eq:trans} on $n$ is different for both polarizations, we can vary the base refractive index until the average radius for each polarization is the same. The final values of the diameter for the analyzed modes are shown in Fig.~\ref{fig:radii}d) and f). The final values for the fiber parameters are displayed in Table \ref{tab:values}.

\begin{table}[htbp]
    \centering
    \begin{tabular}{ccc}
         & $80 \mu$m fiber & $125 \mu$m fiber \\ \hline\hline
        Diameter ($\mu$m) & $77.668 \pm 0.002$ & $122.935 \pm 0.003$ \\ \hline
        $n$ & $1.48723 \pm 0.00002$ & $1.46575 \pm 0.00002$ \\ \hline
        $\dd n/\dd\lambda$ (nm$^{-1}$) & $(-1.47 \pm 0.03)\times 10^{-5}$ & $(-1.34\pm 0.03 )\times 10^{-5}$ \\ \hline
    \end{tabular}
    \caption{Final values of the parameters obtained from the analysis for both fibers. Calculations of the uncertainties can be found in the supplementary information.}
    \label{tab:values}
\end{table}

\section{Outlook}
One of the limiting factors of this method is the reliance on the resonator being circular. The method could be improved to handle fibers with significant eccentricity by modeling the fiber as a cylinder with an elliptical cross section, replacing the Bessel equation with the Mathieu equation and use a new transcendental ``characteristic" equation\cite{yeh2008essence} This would add an extra degree of freedom that would need to be determined to further minimize the standard error in these values.

When WGMs are used for sensing, the changes in the physical environment will often cause a shift in the wavelength due to due to a combination of changing radius and refractive index\cite{foreman2015whispering}. Since by looking at a single mode we cannot differentiate between changes in radius and refractive index, this often means only one physical property can be sensed at a time. It is possible to measure more than one property by using more than one mode, such as changes in refractive index and temperature using two modes with orthogonal polarization\cite{Liu:16}, through a sensitivity matrix. The method introduced here could be used to generalize this to a larger number of properties monitored through the same measurements.

\section{Conclusion}
In this work we developed a non-destructive method to determine the mode number, refractive index, and radius of a cylindrical resonator with a circular cross-section based on spectral measurements of many resonant modes with two orthogonal polarizations. We demonstrated this method using two different fibers that differed in their size and material. The fiber diameters were determined with an accuracy of 2 nm, and the refractive indices to an accuracy of 0.00002. The accuracy of the measurement was limited by the modeling of the fiber as a circle and suggested methods that could improve on this work and how it could be used to improve sensing.

\begin{backmatter}
\bmsection{Funding}
\noindent This work was funded with the support of National Science and Engineering Research Council of Canada (NSERC) through the Discovery Grant program, and the Centre d'Optique, Photonique et Laser (COPL).
\bmsection{Disclosures}
\noindent The authors declare no conflicts of interest.
\bmsection{Data availability}
Data underlying the results presented in this paper are not publicly available at this time but may be obtained from the authors upon reasonable request.
\bmsection{Supplemental document}
See Supplement 1 for supporting content. 
\end{backmatter}

\bibliography{Measurement_of_absolute_radius_refractive_index_and_dispersion_of_a_long_cylinder}

\begin{thebibliography}{10}
\newcommand{\enquote}[1]{``#1''}

\bibitem{fang2012fundamentals}
Z.~Fang, K.~Chin, R.~Qu, and H.~Cai, \emph{Fundamentals of optical fiber
  sensors}, vol. 226 (John Wiley \& Sons, 2012).

\bibitem{s140405823}
J.~Lou, Y.~Wang, and L.~Tong, \enquote{Microfiber optical sensors: A review,}
  {\protect\JournalTitle{Sensors}} \textbf{14}, 5823--5844 (2014).

\bibitem{agrawal2012fiber}
G.~P. Agrawal, \enquote{Optical fibers,} in \emph{Fiber-optic communication
  systems,}  (John Wiley \& Sons, 2012), pp. 24--74.

\bibitem{savchenkov2004kilohertz}
A.~A. Savchenkov, V.~S. Ilchenko, A.~B. Matsko, and L.~Maleki,
  \enquote{Kilohertz optical resonances in dielectric crystal cavities,}
  {\protect\JournalTitle{Physical Review A}} \textbf{70}, 051804 (2004).

\bibitem{yang2016high}
Y.~Yang, S.~Saurabh, J.~M. Ward, and S.~{Nic Chormaic}, \enquote{High-q,
  ultrathin-walled microbubble resonator for aerostatic pressure sensing,}
  {\protect\JournalTitle{Optics express}} \textbf{24}, 294--299 (2016).

\bibitem{dong2009fabrication}
C.-H. Dong, L.~He, Y.-F. Xiao, V.~Gaddam, S.~Ozdemir, Z.-F. Han, G.-C. Guo, and
  L.~Yang, \enquote{Fabrication of high-q polydimethylsiloxane optical
  microspheres for thermal sensing,} {\protect\JournalTitle{Applied physics
  letters}} \textbf{94}, 231119 (2009).

\bibitem{bogatyrev2002change}
V.~A. Bogatyrev, V.~I. Vovchenko, I.~K. Krasyuk, V.~A. Oboev, A.~Y. Semenov,
  V.~A. Sychugov, and V.~P. Torchigin, \enquote{Change in the spectrum of
  optical whispering-gallery modes in a quasi-cylindrical microresonator caused
  by an acoustic pressure pulse,} {\protect\JournalTitle{Quantum electronics}}
  \textbf{32}, 471 (2002).

\bibitem{liang2017resonant}
W.~Liang, V.~S. Ilchenko, A.~A. Savchenkov, E.~Dale, D.~Eliyahu, A.~B. Matsko,
  and L.~Maleki, \enquote{Resonant microphotonic gyroscope,}
  {\protect\JournalTitle{Optica}} \textbf{4}, 114--117 (2017).

\bibitem{vollmer2002protein}
F.~Vollmer, D.~Braun, A.~Libchaber, M.~Khoshsima, I.~Teraoka, and S.~Arnold,
  \enquote{Protein detection by optical shift of a resonant microcavity,}
  {\protect\JournalTitle{Applied physics letters}} \textbf{80}, 4057--4059
  (2002).

\bibitem{ksendzov2004integrated}
A.~Ksendzov, M.~Homer, and A.~Manfreda, \enquote{Integrated optics
  ring-resonator chemical sensor with polymer transduction layer,}
  {\protect\JournalTitle{Electronics Letters}} \textbf{40}, 63--65 (2004).

\bibitem{jiang2020whispering}
X.~Jiang, A.~J. Qavi, S.~H. Huang, and L.~Yang, \enquote{Whispering-gallery
  sensors,} {\protect\JournalTitle{Matter}} \textbf{3}, 371--392 (2020).

\bibitem{del2007optical}
P.~Del’Haye, A.~Schliesser, O.~Arcizet, T.~Wilken, R.~Holzwarth, and T.~J.
  Kippenberg, \enquote{Optical frequency comb generation from a monolithic
  microresonator,} {\protect\JournalTitle{Nature}} \textbf{450}, 1214--1217
  (2007).

\bibitem{arnold2009whispering}
S.~Arnold, D.~Keng, S.~Shopova, S.~Holler, W.~Zurawsky, and F.~Vollmer,
  \enquote{Whispering gallery mode carousel--a photonic mechanism for enhanced
  nanoparticle detection in biosensing,} {\protect\JournalTitle{Optics
  Express}} \textbf{17}, 6230--6238 (2009).

\bibitem{spillane2002ultralow}
S.~Spillane, T.~Kippenberg, and K.~Vahala, \enquote{Ultralow-threshold raman
  laser using a spherical dielectric microcavity,}
  {\protect\JournalTitle{Nature}} \textbf{415}, 621--623 (2002).

\bibitem{grudinin2009brillouin}
I.~S. Grudinin, A.~B. Matsko, and L.~Maleki, \enquote{Brillouin lasing with a
  {CaF\textsubscript{2}} whispering gallery mode resonator,}
  {\protect\JournalTitle{Physical review letters}} \textbf{102}, 043902 (2009).

\bibitem{stewart1982optical}
W.~Stewart, \enquote{Optical fiber and preform profiling technology,}
  {\protect\JournalTitle{IEEE Transactions on Microwave Theory and Techniques}}
  \textbf{30}, 1439--1454 (1982).

\bibitem{ghosh1997sellmeier}
G.~Ghosh, \enquote{Sellmeier coefficients and dispersion of thermo-optic
  coefficients for some optical glasses,} {\protect\JournalTitle{Applied
  optics}} \textbf{36}, 1540--1546 (1997).

\bibitem{birks2000high}
T.~Birks, J.~Knight, and T.~Dimmick, \enquote{High-resolution measurement of
  the fiber diameter variations using whispering gallery modes and no optical
  alignment,} {\protect\JournalTitle{IEEE Photonics Technology Letters}}
  \textbf{12}, 182--183 (2000).

\bibitem{sumetsky2010radius}
M.~Sumetsky and Y.~Dulashko, \enquote{Radius variation of optical fibers with
  angstrom accuracy,} {\protect\JournalTitle{Optics letters}} \textbf{35},
  4006--4008 (2010).

\bibitem{preston2015determining}
T.~C. Preston and J.~P. Reid, \enquote{Determining the size and refractive
  index of microspheres using the mode assignments from mie resonances,}
  {\protect\JournalTitle{JOSA A}} \textbf{32}, 2210--2217 (2015).

\bibitem{ward2014contributed}
J.~Ward, A.~Maimaiti, V.~H. Le, and S.~{Nic Chormaic}, \enquote{Contributed
  review: Optical micro-and nanofiber pulling rig,}
  {\protect\JournalTitle{Review of Scientific Instruments}} \textbf{85}, 111501
  (2014).

\bibitem{BianucciJAP09}
P.~Bianucci, J.~Rodriguez, C.~Clements, J.~Veinot, and A.~Meldrum,
  \enquote{Silicon nanocrystal luminescence coupled to whispering gallery modes
  in optical fibers,} {\protect\JournalTitle{Journal of Applied Physics}}
  \textbf{105}, 023108 (2009).

\bibitem{schott2019}
Schott, \emph{Duran} (2017). Datasheet {PT\_TTS\_1019 GB}.

\bibitem{corning_leaf}
Corning, \emph{Corning LEAF Optical FiberPro} (2019). Datasheet {PI1107}.

\bibitem{stephens1954refractive}
R.~E. Stephens and W.~S. Rodney, \enquote{Refractive indices of five selected
  optical glasses,} {\protect\JournalTitle{Journal of Research of the National
  Bureau of Standards}} \textbf{52}, 303 (1954).

\bibitem{yeh2008essence}
C.~Yeh and F.~I. Shimabukuro, \emph{The essence of dielectric waveguides}
  (Springer, 2008).

\bibitem{foreman2015whispering}
M.~R. Foreman, J.~D. Swaim, and F.~Vollmer, \enquote{Whispering gallery mode
  sensors,} {\protect\JournalTitle{Advances in optics and photonics}}
  \textbf{7}, 168--240 (2015).

\bibitem{Liu:16}
P.~Liu and Y.~Shi, \enquote{Simultaneous measurement of refractive index and
  temperature using a dual polarization ring,} {\protect\JournalTitle{Appl.
  Opt.}} \textbf{55}, 3537--3541 (2016).

\end{thebibliography}

\end{document}


\maketitle

\section{Spectra with labeled mode numbers}
\begin{figure}[h]
    \centering
    \includegraphics[width=\textwidth]{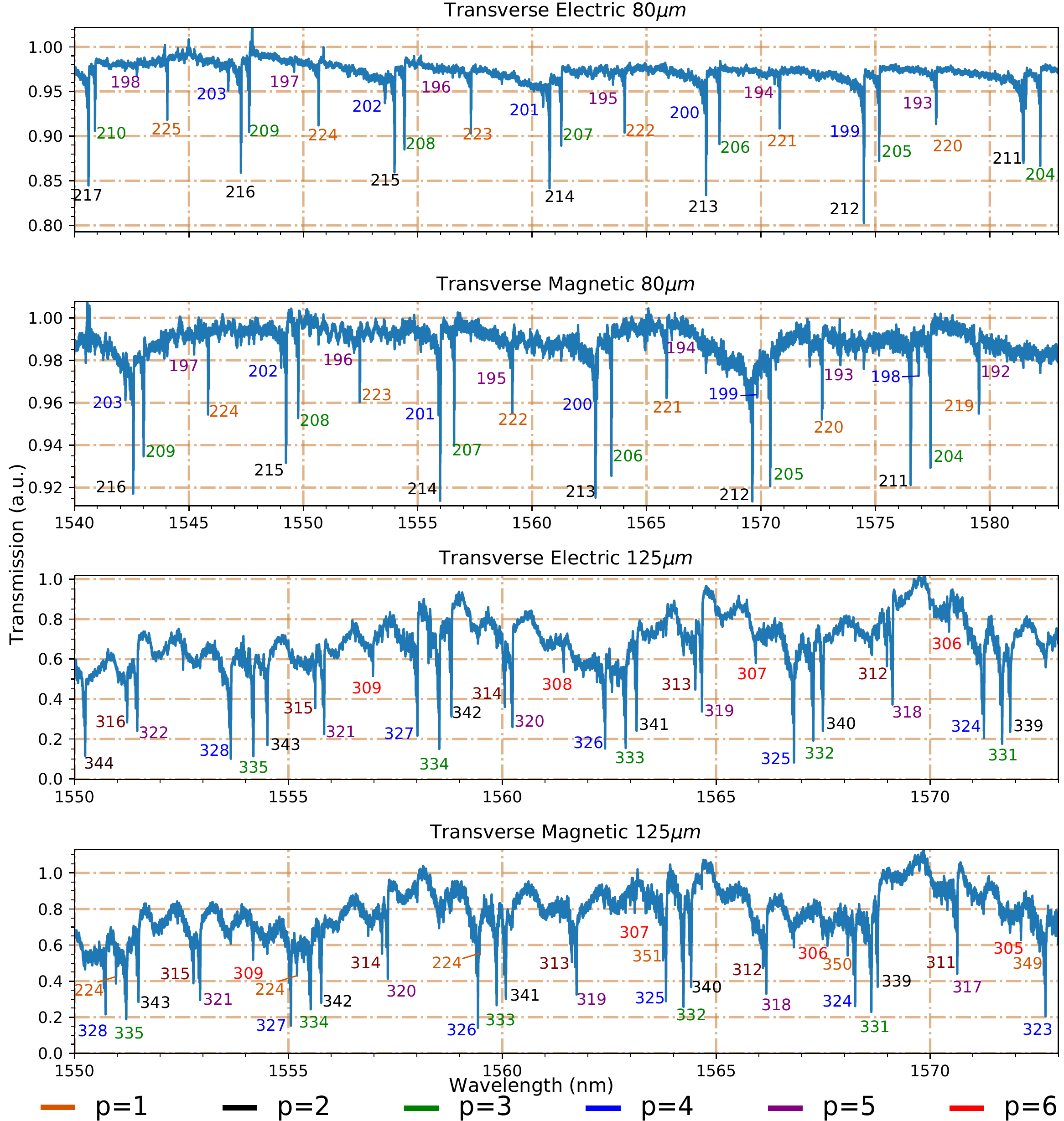}
    \caption{Spectra of both polarizations for both fibers use in the main text with radial mode numbers indicated by color and azimuthal mode number $m$ labeled next to the corresponding modes.}
    \label{fig:my_label}
\end{figure}

\section{Uncertainty calculations}

Since all modes are in the same fiber, they should correspond to the same fiber diameter. Thus, he have taken all the values obtained from different modes for one fiber, and considered the standard error of this set as the uncertainty in the measurement.

Once the uncertainty in the diameter is established, we can propagate this uncertainty to the refractive index value, using standard error propagation, 
\begin{equation}
    \Delta n = \frac{\partial n}{\partial D}\Delta D
\end{equation}
where the $\frac{\partial n}{\partial D}$ coefficent can be numerically calculated from the solutions of Eq. (1) in the main text for each fiber.

The uncertainty for the dispersion is the standard error for the values obtained for each set of modes with common $p$. In other words, we varied the dispersion to minimize the standard error in $D$ for the modes with a common value of $p$ and repeated this for each value of $p$. Then calculated the standard error from this set of data. The uncertainties found from this approach are shown in Table \ref{tab:values_stderr}.

\begin{table}[htbp]
    \centering
    \begin{tabular}{ccc}
         & $80 \mu$m fiber & $125 \mu$m fiber \\ \hline\hline
        $\delta$ D ($\mu$m) & $0.002$ & $0.001$ \\ \hline
        $\delta n$ & $2\times 10^{-5}$ & $1\times 10^{-5}$ \\ \hline
        $\delta (\dd n/\dd\lambda) (nm^{-1})$ & $3.00\times 10^{-7}$ & $3.00\times 10^{-7}$ \\ \hline
    \end{tabular}
    \caption{Uncertainties found via statistical analysis of the mode results.}
    \label{tab:values_stderr}
\end{table}

\subsection{Error propagation}

There are two experimental sources of error in our measurement. The first one is due to accuracy in determining the exact resonant wavelength, which is limited by the wavelength locker at $20pm$. The second is from the deviation of the fiber from a perfectly circular cross section. While it is not straightforward to quantify the effects of the deviation from perfect circularity, we assume it is not large. We can account for the effect of the wavelength uncertainty by doing the corresponding error propagations. Thus, the uncertainty in the radius is can be calculated using 
\begin{equation}
    \delta D = \frac{\partial D}{\partial\lambda}\delta\lambda + \frac{\partial D}{\partial n}\delta n.
    \label{eq:diameter_uncertainty}
\end{equation}
where $\delta D$, $\delta\lambda$ and $\delta n$ are the uncertainty in the diameter, wavelength and refractive index, respectively. The two coefficients can be computed numerically using the solutions of Eq. (1) for the respective fibers.

Since our uncertainty is from the wavelength measurement, the uncertainty in the refractive index can be expressed as a function of the uncertainty of the wavelength uncertainty
\begin{equation}
    \delta n = \frac{\partial n}{\partial\lambda}\delta\lambda,
    \label{eq:n_uncertainty}
\end{equation}
where the coefficient can again be calculated numerically. This leads to
\begin{equation}
    \delta D = \left[\frac{\partial D}{\partial\lambda} + \frac{\partial D}{\partial n}\frac{\partial n}{\partial\lambda}\right]\delta\lambda.
\end{equation}
By cancelling the $\partial n$ we see that both terms are equal.

The uncertainty in the chromatic dispersion due the wavelength uncertainty can be computed using
\begin{equation}
    \delta\frac{\Delta n}{\Delta\lambda} = \frac{\partial}{\partial\lambda}\frac{\Delta n}{\Delta\lambda}\delta\lambda.
\end{equation}
The numerator is the difference in the refractive index of two different modes and the denominator is the difference in wavelength of two different modes. By the linearity of derivatives and the product rule, this can be rewritten as
\begin{equation}
    \delta\frac{\Delta n}{\Delta\lambda} = \left[\frac{(\partial n_m/\partial\lambda) -(\partial n_{m+\Delta m}/\partial\lambda)}{\lambda_m -\lambda_{m+\Delta m}} -\frac{n_m-n_{m+\Delta m}}{(\lambda_m -\lambda_{m+\Delta m})^2}\right]\delta\lambda,
\end{equation}
where the numerator of the first term is computed using Eq.~\ref{eq:n_uncertainty} for the two modes. In this case, the second term dominates. It is worth pointing out that since the terms in the numerator are almost identical in magnitude, it is not important to distinguish between mode number when calculating the uncertainty of $\delta m$ but is important for calculating the uncertainty in the dispersion.
\begin{table}[htbp]
    \centering
    \begin{tabular}{ccc}
         & $80 \mu$m fiber & $125 \mu$m fiber \\ \hline\hline
        $\delta$ D ($\mu$m) & $0.002$ & $0.003$ \\ \hline
        $\delta n$ & $2\times 10^{-5}$ & $2\times 10^{-5}$ \\ \hline
        $\delta (\dd n/\dd\lambda) (nm^{-1})$ & $6\times 10^{-9}$ & $2\times 10^{-8}$ \\ \hline
    \end{tabular}
    \caption{Uncertainties found by propagation of the measured wavelength uncertainty.}
    \label{tab:values}
\end{table}

All the uncertainties due to the wavelength measurement error are given in Table~\ref{tab:values}. 
We see the standard error for the diameter and refractive index are close to the uncertainties due to the wavelength measurement. The standard error for the chromatic dispersion seems larger than the propagated uncertainty due to the wavelength uncertainty. Since the measured chromatic dispersion seems to have a small dependence on the radial mode number, it is possible this discrepancy is due non-circularity or non-homogeneity.

In the main text, we have used the largest of the values shown in Tables \ref{tab:values_stderr} and \ref{tab:values} as the uncertainty for each parameter.
